\documentclass[preprint]{elsarticle}

\usepackage{hyperref}
\usepackage{amsmath,amssymb}
\usepackage{siunitx}
\usepackage{caption}
\usepackage{subcaption}
\usepackage{KITcolors}
\usepackage[version=4]{mhchem}
\usepackage{fourier} 
\usepackage{charter}

\usepackage{soul}

\biboptions{sort&compress}
\newcommand{\sss}{\boldsymbol{s}}
\newcommand{\uu}{\boldsymbol{u}}
\newcommand{\xx}{\boldsymbol{x}}

\newcommand{\dd}{\textrm{d}}
\newcommand{\mua}{\mu_{\textrm{a}}}
\newcommand{\mus}{\mu_{\textrm{s}}}
\newcommand{\thetar}{\theta_\textrm{r}}
\newcommand{\nrel}{\textrm{n}_{\textrm{rel}}}

\makeatletter
\def\ps@pprintTitle{%
    \let\@oddhead\@empty
    \let\@evenhead\@empty
    \def\@oddfoot{\footnotesize\itshape
         {Submitted preprint} \hfill\today}%
    \let\@evenfoot\@oddfoot
    }
\makeatother

\begin{document}

\begin{frontmatter}

\title{Comprehensive computational model for coupled fluid flow, mass transfer and light supply in tubular photobioreactors equipped with glass sponges}

\author[MVM,LBRG]{Albert Mink\corref{IKE}}
\cortext[IKE]{Corresponding author}
\ead{albert.mink@kit.edu}
\author[BVT]{Kira Schediwy}
\author[BVT]{Clemens Posten}
\author[MVM]{Hermann Nirschl}
\author[MATH,LBRG]{Stephan Simonis}
\author[MVM,LBRG,MATH]{Mathias J. Krause}

\address[MVM]{Institute for Mechanical Process Engineering and Mechanics, Karlsruhe Institute of Technology, Germany}
\address[BVT]{Institute of Process Engineering in Life Sciences, Karlsruhe Institute of Technology, Germany}
\address[LBRG]{Lattice Boltzmann Research Group, Karlsruhe Institute of Technology, Germany}
\address[MATH]{Institute for Applied and Numerical Mathematics, Karlsruhe Institute of Technology, Germany}

% ...

\begin{abstract}
The design and optimization of photobioreactors (PBR) can benefit from the development of robust and yet quantitatively accurate computational models, that incorporate the complex interplay of fundamental phenomena. 
At a minimum, the simulation model requires at least three submodels for hydrodynamic, light supply and biomass kinetics as pointed out by various review articles on computational fluid flow models for PBR design. 
By modeling the hydrodynamics, the light-dark-cycles can be detected and the mixing characteristic of the flow besides its mass transport is analyzed. 
The radiative transport model is deployed to predict the local light intensities according to wavelength of the light and scattering characteristic of the culture. 
The third submodel implements the biomass growth kinetic, by coupling the local light intensities to hydrodynamic information of CO2 concentration, to predict the algal growth.

The developed mesoscopic simulation model is applied to a tubular PBR with transparent walls and internal sponge structure. 
The complex hydrodynamics and the homogeneous illumination in such reactors is very promising for CFD based optimization.
\end{abstract}


\end{frontmatter}


\section{Introduction}
%
%
Microalgae pose a renewable and (cost) alternative industrial feedstock by the establishment of photobioreactors (PBR) with optimized light characteristics.
Along the light path algae become more growth limited by insufficient light supply due to the algal absorption and scattering behavior.
In contrast, algae close to the reactor surface are subjected to light saturation or even photoinhibition.
These effects increase the relative proportion of respiration and heat dissipation, respectively. 
This loss of energy reduces biomass growth and is not favorable for industrial processes. Common approaches to overcome (steep) light gradients are internal illuminations~\cite{cornet:10}, the straightforward reduction of the light path using more reactor units, the enlargement of the reactor surface or the installation of light diluting structures in the reactor ~\cite{jacobi1:12}.
All named approaches aim for the enhancement of the surface to volume ratio ~\cite{nwoba:19}.
The increased ratio causes lower light intensities on the reactor surface and (on average) shorter light paths, thus, yields to more homogeneous light supply in the algal culture.

From an engineering perspective, the demand for a high surface to volume ratio can be covered by closed PBR due to their hardly limited light setups. The increased complexity in the light pattern cannot be met by the popular Lambert--Beer-based simulation models.
Instead, new powerful simulation tools are needed, that consider absorption and scattering to predict general light transport in volume and help to speed up the design of novel, more efficient and low cost PBR.

In addition to the improved light pattern in the reactor higher mixing rates  cause the algae to experience more similar conditions in the reactor. The frequent transfer between light zones can be regarded as a homogenization of light conditions and subsequently of the intracellular metabolic reactions. Optimized flows along the light gradient can be initiated by e.g. wall turbulence promoters. 

In CFD simulations, turbulence promoters implemented as profiles at the inner tube wall established better mixing behavior at lower flow velocities, yielding a significantly reduced energy demand~\cite{gomez:15}. Adopting the principle of static mixers, the cultivation efficiency in flat-plate PBR has been increased by optimizing the flow pattern using CFD simulations~\cite{huang:15}.
Some novel reactor designs as the Taylor vortex PBR also show complex flow patterns to lead to promising mixing behaviour~\cite{gao1:15}. The named simulations share as the main characteristic the usage of algae trajectories to evaluate mixing behavior.

Simulation tools for both, light supply and algae trajectory in PBR are advanced and already validated for many algorithms including DOM~\cite{kong:14}, MC~\cite{csogor:01,dauchet:13} and FVM~\cite{huang:11} for light supply, and various Euler--Lagrange models for the trajectories~\cite{nochta:07, gomez:15,lapin:04}.
However, the combination of both tools, eventually taking additional mass transfer into account, remains a numerical challenge.
This often leads to simplified and inaccurate light predictions, considering algebraic light models based on Lambert--Beer and others~\cite{gao:17,huang:15,cornet:09}.

\begin{figure}
\centering
\includegraphics[width=0.8\linewidth]{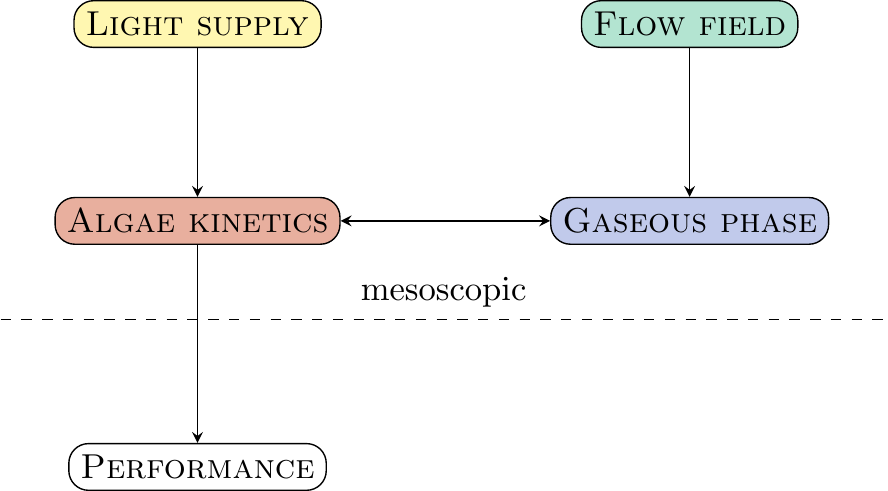}
\caption{Interplay flow field, light supply and mass transfer.}
\label{fig:outline}
\end{figure}

Inaccuracy of PBR simulations limits the progress in reactor design and process development.
However, we need this progress to cover the increasing demand of bio-based resources.
Consistent numerical frameworks that model, at least the light supply, hydrodynamics and biomass growth kinetics, may play an important role in the development of accurate and reliable predictions, see Fig.~\ref{fig:outline}.
Further, the numerical approach renounces complicated coupling and time consuming data interpolation.
Numerical tools might also speed up the prototyping of novel and complex reactor designs as well as analyzing the interplay of flow field, light supply and mass transfer to determine the reactor performance.

This work proposes a consistent three dimensional simulation model for light supply, flow pattern and mass transfer using a numerical framework, based on Lattice Boltzmann Methods (LBM).
The model is applied on a tubular PBR equipped with transparent sponges.
In previous work~\cite{mink:16,mink:18,mink:20} the light transport simulation based on LBM was validated and applied to predict light distribution in PBR.
LBM simulations are a CFD tool with particular strength in multiphysics applications~\cite{trunk:16,maier:16,gaedtke:19,nathen:13} being a suitable framework for the development of a comprehensive computational model for the computation of algae trajectories (Lagrange--Lagrange) and mass transfer (Euler--Lagrange).

\section{Methods}
\label{sec:methods}

\subsection{Description of the glass sponge PBR}
\begin{figure}
\centering
\includegraphics[width=0.8\linewidth]{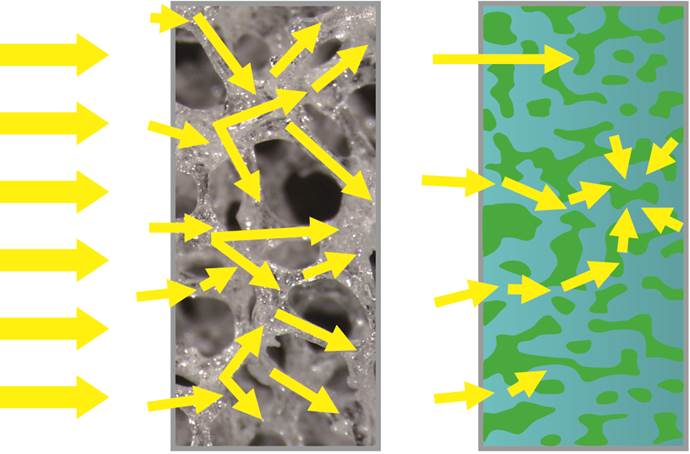}
\caption{Photograph (left) of a transparent sponge with indicated light transmission and scheme (right) of the resulting focus effect due to illumination of the volume in the pores from all direction.
Image from~\cite{jacobi1:12}.}
\label{fig:spongeSketch}
\end{figure}

\begin{figure}
\centering
\includegraphics[width=0.8\linewidth]{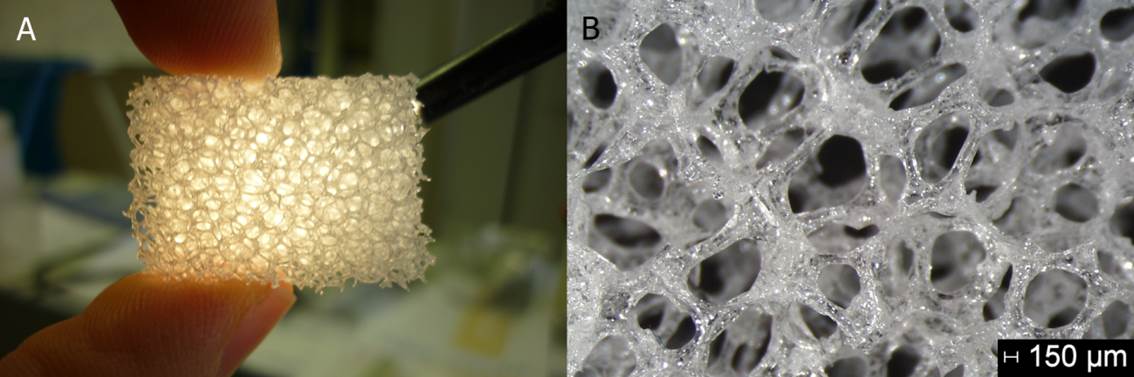}
\caption{Photograph (A) and high resolution image (B) of a transparent sponge show deep and homogeneous light transmission in the glass structure.
Image from~\cite{jacobi1:12}.}
\label{fig:spongePhoto}
\end{figure}

One way to improve the light pattern in microalgal cultures is the installation of transparent polyurethane sponges in a PBR.
Jacobi et al. applied these open-pored sponges manufactured based on the polymer replica technique to flat panel PBR~\cite{jacobi1:12}.
Two major effects qualify the sponges to improve the reactor performance compared to liquid cultures without built-in structures:
(1) The so-called light dilution is achieved by increasing the illumination area by the surface area of the sponges in addition to the surface area of the PBR itself.
The sponges conduct light to deeper positions (Fig.~\ref{fig:spongeSketch}) in the PBR and also decrease the light path in the microalgal culture.
(2) Algae grow within the pores of the sponges and become illuminated from all directions due to the multiple and complex reflections in the glass sponges.
The illumination from all directions towards the center of each pore focuses the light as shown in Fig.~\ref{fig:spongePhoto}.
This focus effect counteracts the attenuation of light due to scattering and absorption of algae along the light path.
Both effects cause more homogeneous or less extreme light intensities compared to liquid cultures in an analogous geometry without sponges, since the attenuation of light shows less decay within the shorter path.
We now apply the sponges on a tubular PBR.
The dimensions of the simulated geometry is \SI{0.05}{\meter} in diameter and \SI{0.05}{\meter} in length, while the initial flow rate is \SI{0.01}{\meter\per\second}.
To evaluate the effect of the sponges simulation results of tubular PBR with sponges are compared to tubular PBR without build-in structures.  
The characteristics of the sponges are scaled up by factor 10 compared to the above-mentioned study with 8 pores per inch, the porosity of 0.9 and the specific surface area~\SI{568}{\square\meter\per\cubic\meter} due to computational limitations.
The absorption of light in the glass remains below \SI{0.1}{\per\meter}. 

\subsection{Mesoscopic modeling and the lattice Boltzmann method}
A crucial difference of the LBM compared to traditional numerical schemes, such as fvm and dom, is the modeling and simulation scope.
Unlike traditional schemes that solve discretized macroscopic equations, the LBM arises from kinetic theory and the Boltzmann equation, where collision rules are defined on a mesoscopic level~\cite{cercignani:87,babovsky:13}.
The mesoscopic scope is a stochastic abstraction of the purely particle based microscopic level and the law of large numbers ensures asymptotically that the mesoscopic collision rule solves macroscopic (target) equations.
LBM are well-known for for its intuitive modeling, intrinsically parallelizable algorithms, the ability to address complex boundary conditions and multiphysic models~\cite{aidun:10,krueger:17}.

\subsection{Light distribution in PBRs}
In spatially homogeneous medium, light transport is governed by the following radiative transport equation (RTE), which models the loss or radiance~$L$ due to absorption~$\mua$ and the redistribution of radiance by scattering~$\mus$
\begin{equation*}
\frac{1}{c} \, \partial_t L
+
\sss \cdot \nabla L
=
-
\left(\mua+\mus\right) \, L
+
\mus
\int_{\Omega}
  \textrm{p}\left(\sss',\sss\right) \,
  L\left(t,\xx,\sss'\right)
  \dd\Omega'
\;.
\end{equation*}
This equation models the change of mesoscopic intensity \textbf{of light}~$L$, in time along a path element~$\sss$, by algae suspension specific absorption and scattering coefficients and scattering integral with phase function~\textrm{p}~\cite{modest:13,chandrasekhar:60}.
Inside a small control volume at position~$\xx$ the local light intensity~$\Phi$ is obtained by formula
\begin{equation}
    \Phi(t,\xx)
    =
    \int L\left(t,\xx,\sss\right) \dd \Omega
    .
\end{equation}
Note, that the rte is formulated with respect to a certain wavelength~$\lambda$, which is dropped for the sake of simplicity.
In previous work, Mink et al.~\cite{mink:16,mink:20} established the radiative transport LBM (RTLBM) as a numerical tool to solve the general RTE.

Spectral light regimes are typically approximated by the Box model, where rte is solved for several wavelengths and the computed intensities are averaged~\cite{huang:12,mcHardy:18}.
This extension might be of interest when improving the light utilization efficiency, since not all absorbed light in the visible spectrum contributes equally to the conversion to biomass. 
Inhomogeneous suspensions are modeled by spacial varying optical parameters~$\mua$ and~$\mus$. The parameter values rise with locally increased biomass concentration and vice versa.
In addition, light sources of arbitrary shape and number can be added by introducing corresponding source terms in the Box model.

The optical characterization of algae suspension with respect to scattering and absorption properties is a non-trivial experimental task.
Measurements depend on the wavelength and growth conditions. Also the scattering must be analyzed within thin suspensions to avoid multiple scattering events that would blur the analysis of one single scattering event~\cite{berberoglu:09,kandilian2:16,mcHardy1:18}.

\begin{table}
\centering
\begin{tabular}{c|c|c}
$\lambda$ in \si{\nano\meter}
& $\hat{\mu}_\textrm{a}$ in \si{\square\meter\per\kilogram}
& $\hat{\mu}_\textrm{s}$ in \si{\square\meter\per\kilogram} \\ \hline
\num{470} & \num{400} & \num{800 }\\
\num{600} & \num{140} & \num{1700}\\
\num{680} & \num{380} & \num{1300}\\
\end{tabular}
\caption{Experimentally obtained absorption and scattering cross-sections of \emph{C. reinhardtii} under optimal growth conditions~\cite{kandilian:16}.}
\label{tab:optical}
\end{table}

Bubbles in the suspension are not considered, since they do not influence the light supply as shown in the experimental and numerical works~\cite{mcHardy1:18,berberoglu:07}.

\subsubsection{Boundary treatment}
\label{ssec:boundary}
The Fresnel's and Snell's law for modeling of reflection and transmission on surfaces yields the following boundary condition.
Radiance that hits a boundary surface from angle~$\theta$ is reflected back and attenuates according to Fresnel's equation.
Only if the incident angle is greater than the critical angle~$\theta_\textrm{c}$, the intensity is totally reflected ($R_F=1$) and there is no transmission.
Otherwise the incident radiation got multiplied by Fresnel's equation
\begin{equation*}
R_F(\theta)
=
\frac{1}{2}
\left(
  \frac{\nrel \cos\thetar -\cos\theta}
       {\nrel \cos\thetar +\cos\theta}
\right)^2
+
\frac{1}{2}
\left(
  \frac{\nrel \cos\theta -\cos\thetar}
       {\nrel \cos\theta +\cos\thetar}
\right)^2, \text{ for } 0\leq \theta \leq \theta_\textrm{c}.
,
\end{equation*}
that results in an attenuation factor due to reflection and transmission.
Parameter~$\nrel$ accounts for the reflection properties of the surface.

In the framework of RTLBM, this boundary condition translates into a partial bounce-back scheme, as shown in previous work~\cite{mink:18}.
Based on reflection properties of the reactor surface, it is determined a physical description of a partial bounce-back coefficient, that models the partial reflection due to Fresnel's equation.

\subsection{Fluid flow regime and Lagrangian particles}
\label{sec:eulerLagrange}
The fluid field is governed by the Navier--Stokes equation for incompressible fluids
\begin{equation*}
\partial_t \uu
+
\left( \uu \cdot \nabla \right) \uu
=
-\frac{\nabla p}{\rho}
+
\nu \nabla^2  \uu
,
\end{equation*}
with fluid velocity~$\uu$, density~$\rho$, pressure~$p$ and kinematic viscosity~$\nu$.
This equation is solved with the open-source LBM implementation presented in~\cite{openLB12}.
To account for the turbulence, a large-eddy simulation model replaces the viscosity with an (local) effective viscosity that models the non-resolved scales~\cite{nathen:13,haussmann:19}.
Basically, the turbulence in the flow regime is either modeled or resolved, depending on the scale, yielding a computational efficient and comprehensive CFD tool.

Individual algae trajectories are computed according to an Euler--Lagrange approach.
Single algae are modeled as Lagrangian point particles of mass~$m_P$ and tracked according to Newton's second law of motion
\begin{equation*}
m_P \partial_t \uu_P = \boldsymbol{F}_P
,
\end{equation*}
for a general force~$\boldsymbol{F}_P$.
Due to the small characteristic size of microalgae (\SIrange{2}{10}{\micro\meter}) the Stokes drag force
\begin{equation*}
\boldsymbol{F}_P
=
\frac{\pi r_p^2}{2} \rho_P C_P (\uu_F -\uu_P)
\end{equation*}
for a drag coefficient~$C_D$ applies, with particle radius~$r_P$ and density~$\rho_P$.
This formula assumes that algae do not affect the fluid flow, which is valid for common biomass concentrations in tubular PBR.
LBM specific implementation details are found in~\cite{henn:16}.

The Euler--Lagrange model allows to study the time evolution of a representative population of algae in the fluid flow and draw conclusion of trapped algae in vortices, the quality of mixing and the induced shear  stress~\cite{lapin:04}.

\subsubsection{The transport of gas}
Mass transport of a species or concentration~$c$ in a flow field $\uu_F$ is modeled by the advection--diffusion equation 
\begin{equation*}
\partial_t c
=
D\Delta c
-\nabla \cdot \uu_F c
+R
\;,
\end{equation*}
also known as Euler--Euler model.
Here, the \ce{CO2} concentration~$c$, due to diffusivity~$D$ and advection by fluid velocity~$\uu_F$.
Sink term~$R$ accounts for the consumption due to photosynthetic conversion, specified later, and is modeled as a linear function of the photo-conversion rate.
This couples the local light intensities with the available gas concentration as discussed later.

The simulation of the present work implements a corresponding Euler--Euler model in the framework of LBM presented and validated in~\cite{trunk:16}.



\subsection{Algae growth model and coupling}
The PBR simulation model couples the local light intensities to the gastransport, by means of a \ce{CO2} consumption according to the (local) rate of photosynthesis.
A simple Monod model, that assumes a light limited biomass growth, is implemented and computes the rate of photosynthesis by
\begin{equation}
P = P_{max} \frac{L}{L_K +L}
\;,
\end{equation}
for maximal photosynthesis rate $P_{max}$ and constant~$L_K$.
A linear connection of specific biomass growth and photosynthesis rate is further assumed.
More complex biomass growth models, e.\,g. accounting for multiple substrate factors and temperature, are known as summarized in~\cite{bechet:13,lee:15,darvehei:18}.
However, the trade off between many empirical parameters and comprehensive modeling has to be managed carefully.

On mesoscopic scale, the local light information is available and the photosynthesis rate is computed easily.
This information is coupled to the \ce{CO2} transport by the sink term~$R$, as mentioned earlier, and represents the gas consumption.
All computational operations are strictly local and renounce the parallel performance of LBM.

\section{Results and discussion}
\label{sec:results}

\subsection{Geometry and computational parameters}
The simulated tubular PBR is of diameter~\SI{0.05}{\meter} and of length~\SI{1}{\meter}, where the sponge-structure is placed at~\SI{0.5}{\meter} from the inlet.
When simulating tubes the hydrodynamics develop after the inlet and are required to settle to avoid numerical errors.
To this aim, the tube is extended artificially to allow the development of the flow pattern.

The discretization in a regular grid of fine and coarse structure at the same time comes with a trade-off of maximal possible grid nodes or lattice cells and the staircase approximation of the fine structure.
Due to memory and computation time limitations, the pore sizes of \SIrange{0.002}{0.004}{\meter} from the original experiments~\cite{jacobi1:12} can not be resolved.
However, the present work considers a pore size of~\SI{0.02}{\meter}, that is resolved by a grid size of~\SI{0.0005}{\meter} yielding an overall number of~\num{15e+6} lattice cells.
Compared to the original pore sizes of~$\approx$\SI{0.003}{\meter} in~\cite{jacobi1:12} the present pore size increased about factor~\num{10}.

\begin{figure}
\centering
\includegraphics[width=\linewidth]{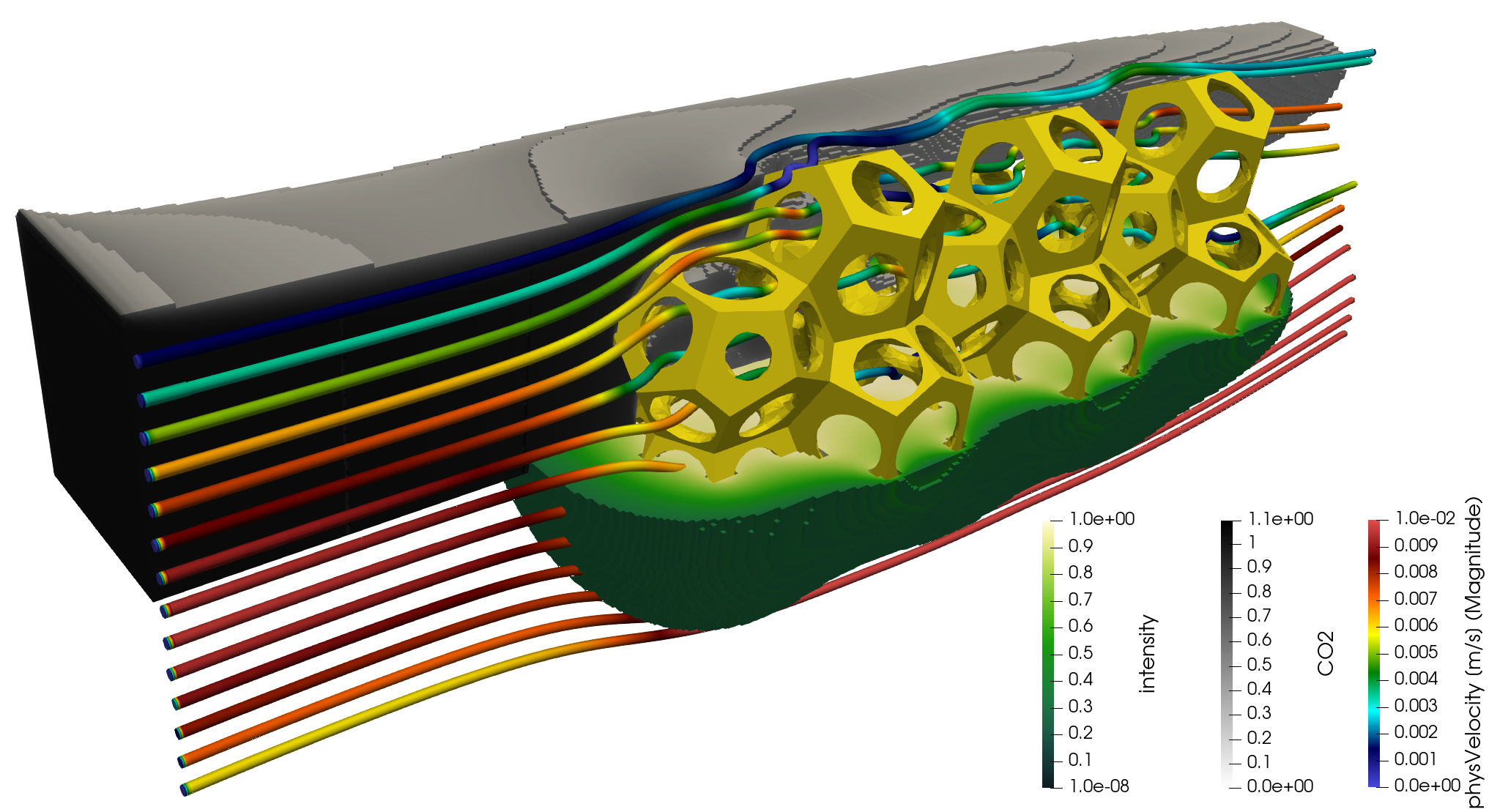}
\caption{Coupled simulation of flow hydrodynamics, gas transport and light distribution in a complex PBR with internal sponge-structure.}
\label{fig:spongePBR}
\end{figure}

Concerning the hydrodynamics, the flow enters with velocity \SI{0.01}{\meter\per\second} in a Poiseuille profile.
The pressure boundary at the end of the channel ensures the outflow, whereas all other walls are modeled as non-slip walls by the LBM bounce-back scheme.
Figure~\ref{fig:spongePBR} shows the geometry of the PBR with the sponge-structure and the simulated flow field, light supply and mass transfer.

\textbf{Simulation sequence:}
First, the light simulation is carried out, where the photo-translucent sponge together with the tubing acts as a radiative source of constant light intensity.
Afterwards, the fluid field is simulated for an Poiseuille inflow profile and non-slip walls with vanishing fluid velocity.
Since the entire geometry is provided by a STL file, the interpolated bounce-back model of Bouzidi et al.~\cite{bouzidi:01} is deployed.
It showed more stability for the staircase approximation of the sponge-structure.
After the full flow pattern is developed, the algae cells are added and tracked accordingly to the Euler--Lagrange approach in Sec.~\ref{sec:eulerLagrange}.

\subsection{Validation of light simulation}
\begin{figure}
\centering
\includegraphics[width=\linewidth]{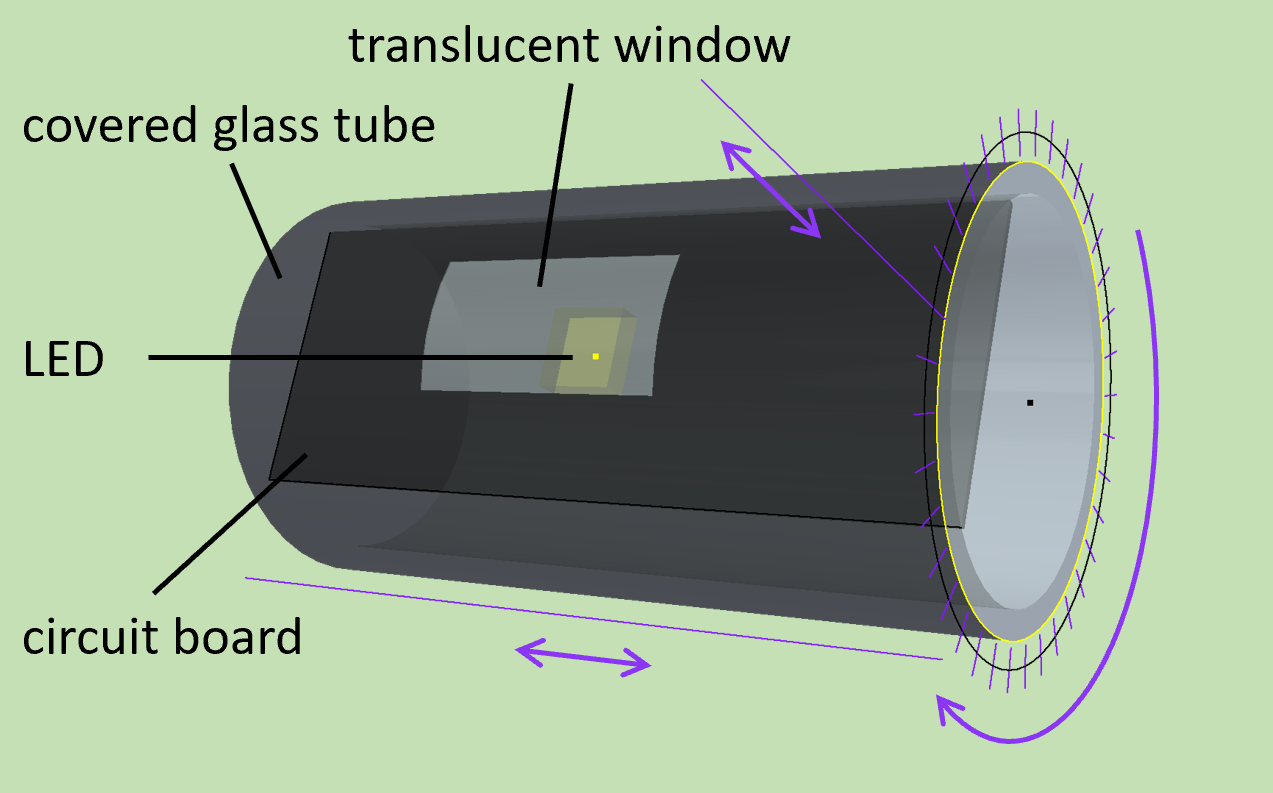}
\caption{}
\label{fig:kiramessung}
\end{figure}

\begin{figure}
\centering
\includegraphics[width=0.6\linewidth]{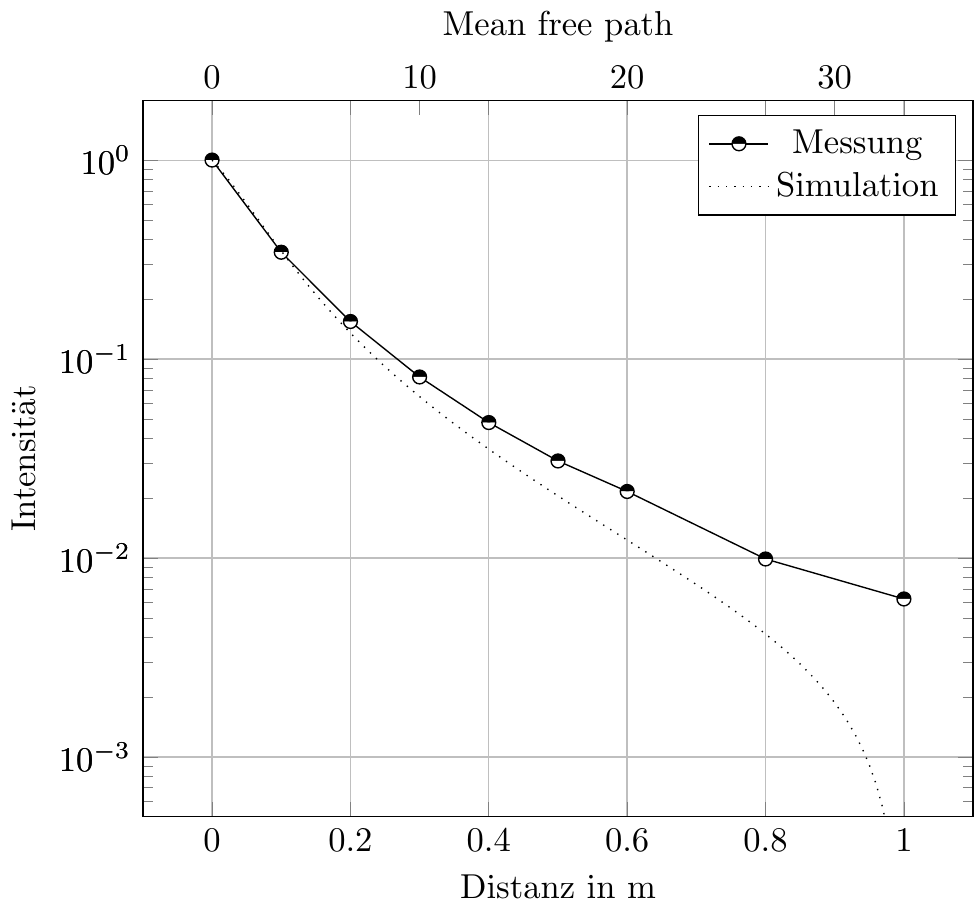}
\caption{Comparison of simulated light intensity with measurements.}
\label{fig:lightVali}
\end{figure}

For validation purposes we measured light intensities in a diluted microalgal suspension.
The low biomass concentration minimized the number of scattering and absorption events to maximize the spatial resolution compared to the sensor's size.
The planar mini quantum sensor (Walz, LS-C, diffuser diameter~\SI{0.003}{\meter}) quantified the photosynthetic active radiation (PAR) emitted by a warm white LED (Nichia, NS6L183BT).
The LED was located in a glass tube to allow light propagation into the suspension. This radial setup causes light gradients along the radius even without algae.
The algal absorption and scattering induces additional attenuation.
To clearly differentiate these effects the glass tube is partially covered to prevent undesired reflection on the glass surface.
For all measurements the planar sensor was oriented towards the center of the long axis of the glass tube.
The measuring configuration allowed shifts in longitudinal direction and variation of the radius up to~\SI{0.05}{\meter} and an angular displacement of \SI{10}{\degree} around the entire glass tube as indicated in Fig.~\ref{fig:kiramessung}.
Measurements and the corresponding three dimensional simulations were performed for a biomass concentration of \SI{9e-5}{\gram\per\cubic\meter}.
The comparison of both data sets along the radius at 0° is exemplarily shown in Figure~\ref{fig:lightVali}.
The agreement of the simulation with measured data is especially convincing above 1\% of the incident light intensity being < \SI{3}{\umol\per\square\meter\per\second}.
Light intensities in this order of magnitude are too low to sufficiently contribute to algal growth or even maintenance requirement of the cells, so should not appear in a pbr and do not substantially impact consecutive growth models.
The light model was, thus, validated and is ready to use for more complex geometries sticking to the same optical principles and characteristics of the cells.


\subsection{Light simulation of complex geometry\textbf{}}

\begin{figure}
\centering
\includegraphics[width=\linewidth]{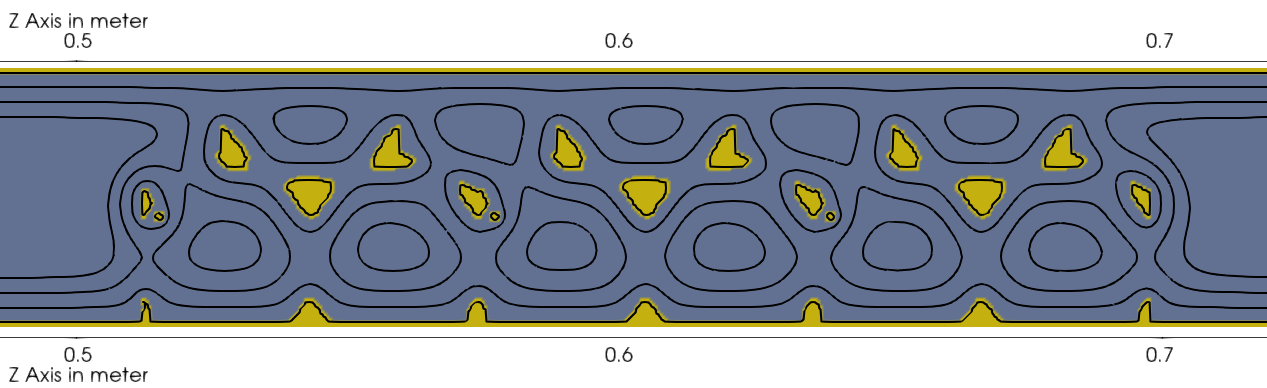}
\caption{Contour lines on the light distribution.
  Yellow parts are light sources and the algae suspension is colored blue.}
\label{fig:spongeRTE}
\end{figure}

Light simulation results of the tubular pbr with and without embedded sponges are depicted on the contour plot in Fig.~\ref{fig:spongeRTE}. To evaluate the general effect of the sponges in the tubular pbr two ideal scenarios are compared: In an ideal theoretical setup a certain light intensity radiates from the pbr wall and the sponges completely into the culture.  The incident light intensity is considered to be independent from the intensity inside the sponge structure to achieve maximum homogenized conditions. (1) The sponges and the tubular surface emit light at equal intensities. (2) The same amount of photons per time as in (1) is emitted by the pbr surface only.  The sponge-structure enlarges the light emitting surface of the pbr. Fig.~\ref{fig:spongeRTE} allows to study the light supply in the classic tube reactor illuminated by its walls as well as the novel reactor with the internal sponge-structure that dilutes the illumination into the suspension.
Without the sponges the steep radial gradient shadows most of the suspension and results in a heterogeneous light distribution with dark zones in the center of the tube. Close to the walls part of the reactor volume is subjected to photo-saturation and results in light limitation in deeper zones.
Increasing the light intensity might reduce the dark zones. However, the higher incident light intensity might expose the suspension close to the light source to photo-inhibition, so induces inefficient light use or even damages the algal cells.
The approach of increasing the light intensities translates the deficit to other reactor regions and barely solves the issue of heterogeneous light supply. Spreading the light sources over a larger surface and into deeper zones of the reactor by the sponges homogenizes the light supply clearly according to the simulations. 

Concluding, the implementation of a sponge-structure inside the pbr is a promising strategy to homogenize the light distribution. The increase of the illuminating surface goes along with a reduction of the incident light intensity inversely proportional to the increase of illumination surface. This so-called light dilution reduces both dark zones and zones of light saturation or even photo-inhibition. The more homogeneous light supply enables higher biomass productivity. This effect is even more substantial in a real setup, since light gradients become steeper, when illumination from the sun occurs from one side of the pbr only.

\subsection{Fluid flow regime and gaseous transport}
\begin{figure}
\centering
\includegraphics[width=\linewidth]{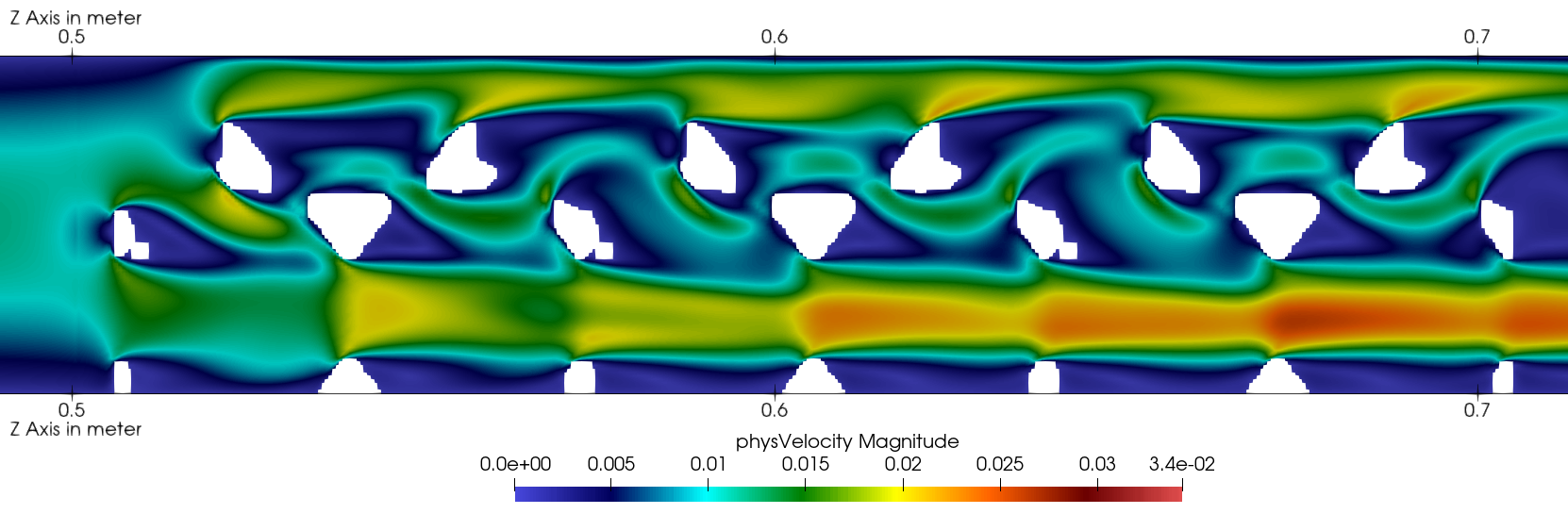}
\caption{Local velocities for an inflow of \SI{0.01}{\meter\per\second}.
  White regions are the sponge-structure modeled as slip walls.}
\label{fig:spongeCFD}
\end{figure}

The sponge-structure attains a complex flow pattern through the channel that enhances the radial mixing.
While the fluid passes the sponge, the local constrictions lead to high velocities and a complex flow pattern develops for inlet velocity as slow as~\SI{0.01}{\meter\per\second}, see Fig.~\ref{fig:spongeCFD}. 

When the flow field is fully developed, the Lagrance particles are seeded randomly at the inlet and are tracked through the pbr. 

\begin{figure}
\centering
\includegraphics[width=0.8\linewidth]{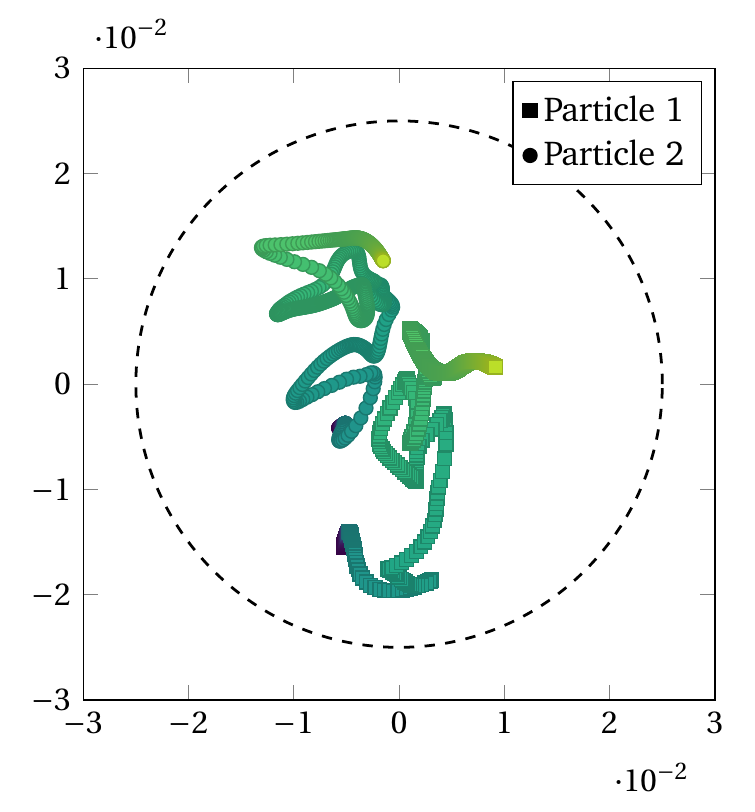}
\caption{Cross section of tubular reactor shows the time evolution of two algae trajectory in xy-plane.
  Trajectory starts is indicated by the dark blue and the end by light green.}
\label{fig:trajectory}
\end{figure}

Figure~\ref{fig:trajectory} depicts the radial position of such tracked algae cells in the reactor.
The color gradient indicates the time evolution in the flow field from the beginning (dark) to the end (light).
Along its trajectory the algae is subjected to radial movements that indicate non-laminar flow pattern and proves the mixing characteristic of this particular reactor design with built-in sponges.
The information of the spacial positions is further analyzed by the locally subjected light intensities that are depicted for the same particles in Fig.~\ref{fig:exposedLight}.
Analyzing the algae light history along the z-direction, the sponge-structure increases the general light supply clearly.
Before and after the sponge the subjected light decreases and remains at a constant level.
The latter shows also the laminar character of the flow field and how algae cells are not mixed and accordingly subjected to different light regimes.

\subsection{Biomass growth}
\begin{figure}
\centering
\includegraphics[width=0.8\linewidth]{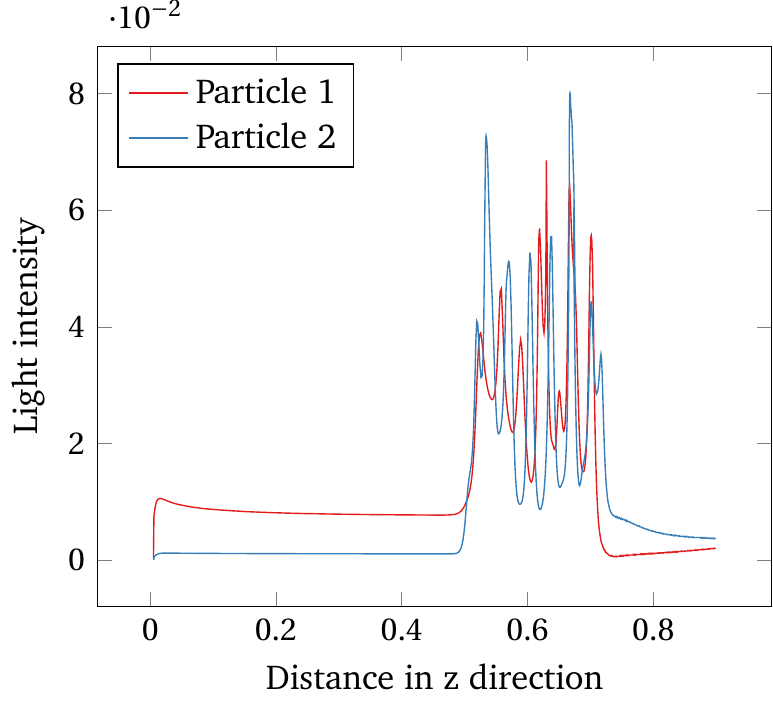}
\caption{Experienced light intensity plotted over traveled z-distance.
  The increase of subjected light intensity at z-position~\SI{0.5}{\meter} is due to the sponge-structure.
  Before and after the sponge-structure, the subjected light is constant.}
\label{fig:exposedLight}
\end{figure}

Figure~\ref{fig:exposedLight} accounts for the light that the algae is exposed to in the flow field.

\section{Conclusion}

In the present work, we propose a consistent three dimensional simulation model for light supply, flow pattern and mass transfer. 
The model makes use of a numerical framework, based on Lattice Boltzmann Methods (LBM) and requires at least three submodels for hydrodynamics, light supply and biomass kinetics, respectively. 
Subsequently to its methodological development, the novel mesoscopic simulation model is successfully applied to a tubular PBR with transparent walls and internal sponge structure. 
In particular, by modeling the hydrodynamics, the light-dark-cycles are detected and the mixing characteristic of the flow besides its mass transport is analyzed. 
The radiative transport model is deployed to predict the local light intensities according to wavelength of the light and scattering characteristic of the culture. 
The third submodel implements the biomass growth kinetic, by coupling the local light intensities to hydrodynamic information of CO2 concentration, to predict the algal growth.

It is a fact that CFD-based methods are very promising for the optimization of the complex hydrodynamics and the homogeneous illumination in the design processes of PBR.
Conclusively, the design as well as the optimization of such reactors benefits directly from the here presented robust and yet quantitatively accurate computational model which incorporates the complex interplay of fundamental phenomena.

\section*{Acknowledgment}
This work was performed on the computational resource ForHLR~I (ForHLR~II) funded by the Ministry of Science, Research and the Arts Baden-W\"urttemberg and DFG ("Deutsche Forschungsgemeinschaft").
This work was supported by the Deutsche Forschungsgemeinschaft (DFG) Grant No. 322739165.



\end{document}